\documentclass[aps,prx,twocolumn,amsmath,amssymb,reprint,superscriptaddress]{revtex4-1}

\usepackage[utf8]{inputenc}
\usepackage{amsmath}
\usepackage{amssymb}
\usepackage{graphicx}
\usepackage{dcolumn}
\usepackage{bm}
\usepackage{makeidx}
\usepackage[alsoload=synchem]{siunitx} 
\usepackage{physics}
\usepackage[colorlinks,allcolors=blue]{hyperref}
\usepackage{cleveref}	
\usepackage[usenames, dvipsnames]{color}

\begin{document}
\title{Quasiparticle poisoning of Majorana qubits}

\author{Torsten Karzig}
\affiliation{Microsoft Quantum, Station Q, Santa Barbara, California
93106, USA}

\author{William S. Cole}
\affiliation{Microsoft Quantum, Station Q, Santa Barbara, California
	93106, USA}

\author{Dmitry I. Pikulin}
\affiliation{Microsoft Quantum, Station Q, Santa Barbara, California
	93106, USA}
\affiliation{Microsoft Quantum, Redmond, Washington 98052, USA}

\date{\today}

\begin{abstract}
Qubits based on Majorana zero modes are a promising path towards topological quantum computing.
Such qubits, though, are susceptible to quasiparticle poisoning which does not have to be small by topological argument.
We study the main sources of the quasiparticle poisoning relevant for realistic devices -- non-equilibrium above-gap quasiparticles and equilibrium localized subgap states.
Depending on the parameters of the system and the architecture of the qubit either of these sources can dominate the qubit decoherence.
However, we find in contrast to naive estimates that in moderately disordered,  floating Majorana islands the quasiparticle poisoning can have timescales exceeding seconds.
\end{abstract}

\maketitle

Majorana zero modes (MZMs) provide a basis for topologically protected qubits \cite{kitaev2003fault, nayak2008non, lutchyn2018majorana}.
The topological protection means that dephasing of the MZM-based qubit (Majorana qubit) is exponentially small in the separation of the MZMs in space and in the energy gap of the separating region.
Both of these quantities can be controlled in experiment which leads to the potential of long dephasing times.

Majorana qubits encode information in the joint parity of MZM pairs.
Dephasing of a Majorana qubit can thus only happen via incoherent exchange of parity between the MZMs \cite{Knapp18} or via processes of uncontrolled exchange of quasiparticles (QPs) between the Majorana subspace and other fermionic modes \cite{goldstein2011,rainis2012majorana,budich2012}.
The latter is called quasiparticle poisoning (QPP) of the Majorana qubit.
QPs may have different origins -- they may be inside the Majorana qubit at above-gap energies excited by temperature or external perturbations, they may come from the environment, or they may be located in non-topological subgap states.
Exponential suppression of QPP in parameters of the system can be shown for a system decoupled from gapless leads under the assumptions of thermal equilibrium and a spectral gap in the system \cite{menard2019suppressing, aseev2019degeneracy}.
In the present work we analyze how breaking these assumptions in real systems affects QPP.

Non-equilibrium QPs are present in any realistic physical system and have been explored extensively in the context of superconducting qubits \cite{Martinis2009, sun2012measurements, riste2013millisecond, Wang2014,vool2014, Riwar16, hosseinkhani2017optimal, Grunhaupt2018, serniak2018hot,Serniak19} and dedicated devices \cite{aumentado2004nonequilibrium, shaw2008kinetics, vanwoerkom2015, hays2018direct, van2018magnetic}.
In a conventional superconductor a single QP cannot relax to the condensate consisting of Cooper pairs.
Once QPs are created by an external perturbation the only way to get rid of them is to pair them up.
This process becomes slow for low densities of the QPs \cite{Martinis2009, Bespalov16}.
Therefore, even for small rates of QP creation, superconductors typically have a non-equilibrium density of QPs that significantly exceeds the expected thermal occupation.

The effects of non-equilibrium QPs on the Majorana qubits can be described by relaxation events of the QPs into the MZMs.
The QPs are described by the density $n_{\text{qp}}$.
In order to determine the lifetime (or dephasing time) of the qubit we have to determine the relaxation rate into MZMs given a certain $n_{\text{qp}}$ and then find an expression for typical QP densities.
The latter can be obtained from steady state solutions of a model with a certain rate of exciting higher-energy QPs, subsequent relaxation and recombination.

The potential danger of QPP for Majorana qubits is widely acknowledged in the literature \cite{goldstein2011, rainis2012majorana,budich2012, Karzig2017, Knapp18} and the timescales for the poisoning influence the design of the Majorana-based quantum computer \cite{knapp2018modeling}.
However, there exist few quantitative estimates for the corresponding decoherence times.
Moreover, the existing estimates \cite{rainis2012majorana,Knapp18} rely on values for $n_{\text{qp}}$ that are typical for conventional superconductors and do not take into account how the presence of the MZM itself will change $n_{\text{qp}}$.

Finally, the effect of disorder and subgap states is mainly discussed in the literature in terms of the effect on the topological transport gap and transition from the topological to the trivial phase due to disorder \cite{Brouwer2011, brouwer2011topological, lobos2012interplay, adagideli2014effects}.
The disorder, however, suppresses the spectral gap faster than the transport gap and thus causes the presence of subgap states well before the topological transition. Such subgap states present a reservoir where parity may leak from the MZMs, thus causing decoherence from equilibrium QPP.

\begin{figure}
	\includegraphics[width=0.95\columnwidth]{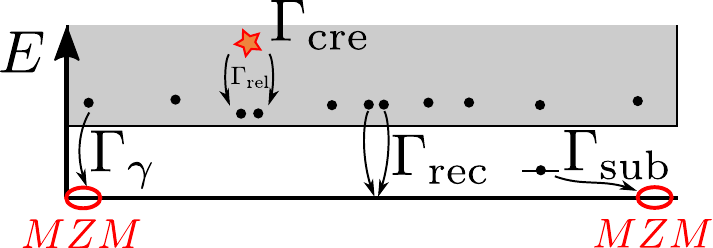}
	\caption{Schematic representation of the relevant quasiparticle processes in a topological superconductor. Above-gap QPs are created at a rate $\Gamma_{\rm cre}$ which leads to a density of non-equilibrium QPs, which quickly relaxes to an energy window close to the gap edge via the rate $\Gamma_{\rm rel}$. After relaxing to the gap edge, the quasiparticles can relax further either by pairwise recombination into Cooper pairs at a rate $\Gamma_{\rm rec}$ or by relaxing into MZMs at rate $\Gamma_\gamma$. At low densities $\Gamma_{\rm rec}$ becomes slow and the QP density close to the MZM is suppressed and dominated by the more effective $\Gamma_{\gamma}$. Another source of quasiparticle poisoning can be due the the presence of low energy subgap states that may be present in the topological superconductor. Leakage of the parity from the MZMs into such state is described by the rate $\Gamma_{\rm sub}$.}
	\label{fig:1}
\end{figure}

The aim of the present paper is to provide a self-consistent estimate of QPP taking into account the peculiarities of mesoscopic topological superconductors including the finite volume of the superconductor and equilibrium QPP due to the presence of subgap states.

{\it Poisoning due to non-equilibrium QPs.---}We start by considering non-equilibrium QPP resulting from relaxation of above-gap non-equilibrium QPs into one of the computational Majoranas with a rate $\Gamma_{\gamma}$. See Fig.~\ref{fig:1} for a schematic representation of the relevant processes.
The non-equilibrium QP density in superconductors can result from a steady rate of QP creation $\Gamma_{\rm cre}=2\gamma_{\rm br}N_{\rm CP}$ via breaking of some of the $N_{\rm CP}$ Cooper pairs.
The exact nature of the Cooper pair breaking process may be due to stray radiation, cosmic rays, etc. but is unimportant for our model.
In accordance with the literature \cite{Martinis2009, Segall04}, we further assume that the resulting high energy QPs relax quickly to the gap edge but only slowly recombine into Cooper pairs by annihilating with another QP with a rate  $\Gamma_{\rm rec}$.
The latter can be estimated by $\Gamma_{\rm rec}=\tilde{\gamma}_0 n_{\rm qp}^2 V/n_{\rm CP}$, where $\tilde{\gamma}_0^{-1}$ is a characteristic time scale of the electron-phonon coupling, $V$ is the total volume of the island, and $n_{\rm CP}$ is the density of Cooper pairs.
The factor $n_{\rm CP}^{-1}$ also describes the average volume of a QP in the energy range of interest, as $n_{\rm CP}=D(E_F)\Delta$ in terms of the density of states $D(E_F)$ and the gap $\Delta$ of the superconductor.
Crucially, $\Gamma_{\rm rec}\propto n_{\rm qp}^2$ becomes slow for small densities which makes relaxation processes into MZMs that become available in topological superconductors highly relevant.

As we detail below, for practical sizes of the superconducting islands, the rate of relaxing into MZMs $\Gamma_\gamma$ (see eq. \eqref{eq:Gamma_gamma}) would be by far the fastest relaxation channel for QPs.
Similar to QP traps based on superconducting vortices  \cite{Ullom98,Peltonen11,Wang2014,Nguyen13,Nsanzineza14,Taupin16} MZMs act as efficient QP traps. The resulting non-equilibrium density of the QPs are lower by several orders of magnitude than what one would expect from the conventional estimates \cite{Martinis2009, Bespalov16} and make them virtually non-existent in some of the proposed Majorana qubits.
We note that the encouraging estimates we obtain below do not transfer directly to the setups based on  bulk superconductors, like \cite{hyart2013flux}.

Following the literature \cite{Segall04, Wang2014, Riwar16, hosseinkhani2017optimal}, we describe the behavior of above-gap QPs by diffusive dynamics of an energy-independent density $n_\text{qp}$ of QPs close to the gap edge. Neglecting the energy dependence is based on the observation that the QP relaxation rate $\Gamma_\text{rel}\gtrsim (10 \text{ns})^{-1}$ \cite{Segall04} is typically by far the fastest of the rates introduced in Fig.~\ref{fig:1} and can thus be eliminated together with the above-gap energy dependence. Within this model we consider MZMs located at positions $\mathbf{x}=\mathbf{x}_i$ and take into account the remaining creation and relaxation processes defined above. The diffusion equation reads
\begin{eqnarray}
\dot{n}_{\rm qp} &=& 2\gamma_{\rm br}n_{\rm CP} +D\nabla^2 n_{\rm qp} \nonumber \\ 
&-& \sum_i\gamma_0 n_{\rm qp} f(\mathbf{x}-\mathbf{x}_i) - \tilde{\gamma}_0 n_{\rm qp}^2/n_{\rm CP} \, . 
\label{eq:diff}
\end{eqnarray}
Here, $\gamma_{\rm br}$ is the rate of breaking Cooper pairs, $D$ is the diffusion constant of the above-gap QPs, $\gamma_0^{-1}$ a characteristic time scale for QP relaxation into a MZM and $f(\mathbf{x})$ describes the local extent of a MZM located around $\mathbf{x}=0$.
While in general $f(\mathbf{x})$ depends non-trivially on the overlap of MZM and QP wavefunctions \cite{Knapp18}, in the limit of a weakly changing QP density \footnote{We confirm that the described system operates in this limit as $L_\gamma \gg \xi$} relative to the topological coherence length we can write $f(\mathbf{x})\approx V_{\rm MZM} \delta(\mathbf{x})$, where $V_{\rm MZM}=(\int d V |\psi_{\rm MZM}|)^2$ is a measure of the volume of the MZM with wavefunction $\psi_{\rm MZM}$.
In an effectively one-dimensional system  $V_{\rm MZM}^{(1d)}=\xi$.

Note that Eq.~\eqref{eq:diff} neglects the hybrid character of typical realizations of topological superconductors as semiconductor-superconductor heterostructures. Although diffusion in the semiconductor can be faster than in the superconductor, the much larger density of states in the superconductor leads to Eq.~\eqref{eq:diff} being an excellent approximation when using the diffusion constant of QPs in the superconductor as long as the regime of very weak superconductor-semiconductor coupling is avoided. For a more detailed discussion see Supplement \ref{sup:coupled}. 

We first review the argument giving the non-equilibrium QP density in an isolated trivial superconductor.
To this end we set $\gamma_0=0$.
The steady state solution of Eq.~(\ref{eq:diff}) then yields a constant density $n_{\rm qp}(x)=n_{\rm SC}$, with
\begin{equation}
n_{\rm SC} = \sqrt{\frac{2\gamma_{\rm br}}{\tilde{\gamma_0}}} n_{\rm CP}\,.
\label{eq:island_steady_state}
\end{equation}
such that $\Gamma_{\rm cre}(n_{\rm SC}) = \Gamma_{\rm rec}(n_{\rm SC})$. 

For the remainder of the manuscript we will use parameters for an Al-proximitized nanowire of total volume $V_{\rm SC} = 10 \mu {\rm m}\times 200 {\rm nm} \times 10 {\rm nm}=2\cdot 10^{-2}\mu {\rm m}^3$ as such system is the closest to practical applications \cite{lutchyn2018majorana}. 
We summarize the parameters in Table \ref{table:params}.
All the estimates are straightforward to perform for other superconductors and host systems \cite{pendharkar2019parity, mi2013proposal, manousakis2017majorana}.
In the Al-based system $n_{\rm cp} = D(E_F)\Delta \approx 3\cdot 10^6/\mu \mathrm{m}^3$.
Using $\gamma_0^{-1},\tilde{\gamma}_0^{-1} \sim 50 $ns \cite{Knapp18} and $n_{\rm SC}=0.01 \dots 10 \mu{\rm m}^{-3}$ \cite{shaw2008kinetics,Martinis2009,vool2014,serniak2018hot,Serniak19,Vepsalainen20} we find $\gamma_{\rm br}^{-1}\sim 10^{4} \dots 10^{10} $s.
Note that this gives quite small total rates of QP creation even when multiplying by the number of Cooper pairs in a typical sample of size $V_{\rm SC}$. Using these numbers we find $\Gamma_{\rm cre}^{-1}\sim 0.1 {\rm s} \dots 1 {\rm day}$. 

Note that the above estimate relies on uniformly distributed completely delocalized QPs and thus provides an upper bound for the creation rate.
As noted e.g. in Ref.~\cite{Bespalov16} the annihilation itself can reduce the probability for QPs to be close enough to recombine especially if they are localized due to disorder.
This effect makes recombination less efficient and leads to estimates with even smaller rates $\Gamma_{\rm cre}$ that would be consistent with the observed QP densities.
Ref.~\cite{Bespalov16} also provided an encouraging estimate for the QP creation rate (per volume) due to cosmic radiation $\sim10^{-4}{\rm s}^{-1} \mu{\rm m}^{-3}$ which corresponds to $\Gamma_{\rm cre}^{-1}\sim 10 \,{\rm days}$.

Let us now consider the rate of relaxation of a density $n_{\rm qp}$ into the localized MZMs given by
\begin{equation}
	\Gamma_\gamma=\gamma_0 n_{\rm qp}V_{\rm MZM}. \label{eq:Gamma_gamma}
\end{equation}
The rate $\Gamma_\gamma$ is also QP poisoning rate of the qubit. A naive estimate of the qubit poisoning rate can be obtained by using typical densities $n_\text{SC}$ of trivial superconductors as an estimate of $n_\text{qp}$ in Eq.~\eqref{eq:Gamma_gamma} \cite{Knapp18}. As we will show below it is crucial that the QP density is determined self-consistently taking into account the presence of MZMs and estimates based on the density in trivial superconductors vastly overestimate the poisoning rate in mesoscopic supercondcutors.

The importance of the relaxation into MZMs in comparison to the bulk QP recombination can already be revealed by examining for a fixed $n_{\rm qp}$ the ratio of $\Gamma_\gamma$ to the rate of pairwise QP annihilation $\Gamma_\gamma/\Gamma_{\rm rec} =n_{\rm cp} V_{\rm MZM}/n_{\rm qp} V$. Using for the estimate of the volume occupied by the Majorana wavefunction $V_{\rm MZM}\sim 200\nm \times 10\nm \times 100\nm = 2 \cdot 10^{-4} \mu{\rm m}^3$, we obtain that at densities similar to the zero field densities of non-equilibrium QPs $n_{\rm qp}\approx n_{\rm SC}$, $\Gamma_\gamma/\Gamma_{\rm rec} \sim 10^2 \dots 10^5\, (1\mu{\rm m}^3/V)$.
With typical device volumes of order $V_{\rm SC}$ we therefore find that the relaxation into MZMs is by far the dominant relaxation process.

\begin{figure}
	\includegraphics[width=0.95\columnwidth]{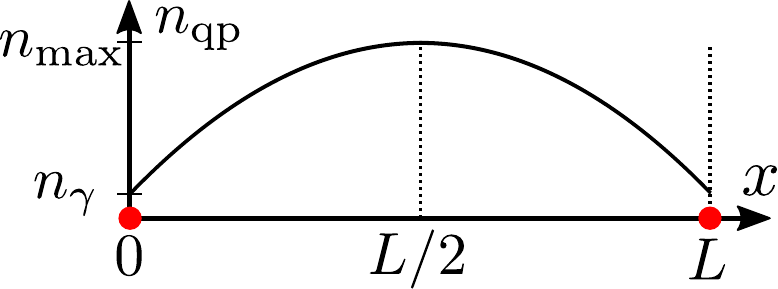}
	\caption{Density profile of non-equilibrium QPs in a quasi one-dimensional system of length $L$ with MZMs located at the ends of the system. Close to the MZMs, the density is suppressed to a value $n_\gamma$ while it reaches $n_{\rm max}$ at the maximal distance from the MZMs.}
	\label{fig:2}
\end{figure}

The above suggests that for topological superconductors of moderate size we can safely neglect the pair recombination in Eq.~\eqref{eq:diff}.
This allows to directly integrate Eq.~\eqref{eq:diff}.
For concreteness we consider in the following a quasi one-dimensional system of length $L$ with two MZMs located at $x=0$ and $x=L$ with $f(x)=\xi \delta(x)$, see Fig.~\ref{fig:2}. Due to the symmetry of the system it is sufficient to focus on the region of $x\in [0, L/2]$. Using boundary conditions of vanishing current $\partial_x n_{\rm qp}(x)=0$ at $x=0$ \footnote{Due to the delta function form of $f(x)$, $\partial_x n_{\rm qp}(x)$ jumps at $x=0$. The boundary condition of vanishing derivative at $x=0$ should be understood at infinitesimally negative $x$.} and $x=L/2$ we obtain the steady state solution of the diffusion equation
\begin{eqnarray}
n_{\rm qp}(x)= \left(1+ \frac{\gamma_0 \xi}{D}x\right)n_\gamma-\frac{\gamma_{\rm br}n_{\rm CP}}{D} x^2\, \label{eq:diff_sol}
\end{eqnarray}
with $n_\gamma=n_{\rm qp}(0)=\gamma_{\rm br}n_{\rm CP}L/ \gamma_0 \xi$. The expression of the QP density at the MZM reflects the balance of the total relaxation and creation rates.

The solution of Eq.\eqref{eq:diff_sol} is depicted in Fig.~\ref{fig:2} and is characterized by a minimum in the QP density $n_{\rm qp}=n_\gamma$ close to the MZMs and a maximal density $n_{\rm qp}=n_{\rm max}$ in-between the MZMs. We now discuss the important length scales of the problem. From Eq.~\eqref{eq:diff_sol} one can extract the length scale $L_\gamma=D/\gamma_0\xi$ over which the density changes only weakly. In a system of size $L<L_\gamma$ the diffusion time to explore the system is shorter than the typical relaxation time. This lead to a homogeneous density, i.e. $n_{\rm max}/ n_\gamma\approx 1$. Using a diffusion constant typical for of Al QPs $D=2 \mu{\rm m}^2{\rm ns}^{-1}$ \cite{Wang2014}, $\xi\approx 200$nm \cite{Albrecht16,Vaitiekenas20}, and $\gamma_0^{-1}\sim 50{\rm ns}$ we obtain $L_\gamma\sim  1$mm. We thus expect that Majorana qubits based on moderately sized islands of topological superconductors \cite{Karzig2017,Plugge17} are in the regime of a small constant density of QPs dominated by relaxation into MZMs. The corresponding QP-limited decoherence times are thus given by
\begin{equation}
	\Gamma_{\gamma,\text{meso}}^{-1}=\Gamma_{\rm cre}^{-1} \sim 0.1 {\rm s} \dots 1 {\rm day}\,.
	\label{eq:meso}
\end{equation}
While the estimate for creation rate of QPs $\Gamma_{\rm cre}$ involves significant uncertainties, it is reasonable to assume that even in the presence of QPP qubit lifetimes exceeding seconds are possible.

\begin{table}
	\begin{tabular}{c | c | c}
		Parameter & Value & Reference
		\\
		\hline\hline
		Volume of superconductor $V_{\rm SC}$ & $2\cdot 10^{-2}\mu$m$^3$ & \cite{lutchyn2018majorana}\\
		Electron-phonon coupling $\gamma_0^{-1}$ & 50ns & \cite{Knapp18} \\
		QP density in bulk SC $n_{\rm SC}$ & $0.01\ldots 10\mu$m$^{-3}$ & \cite{shaw2008kinetics,Martinis2009}, \ldots\\
		Majorana volume $V_{\rm MZM}$ & $2\cdot 10^{-4}\mu$m$^3$ &\cite{lutchyn2018majorana}\\
		Diffusion coefficient in Al $D$ & $2\mu$m$^2$ns$^{-1}$ & \cite{Wang2014}
		
	\end{tabular}
\caption{Parameter values used in our estimations
	\label{table:params}}
\end{table}

To obtain estimates for the case of Majorana qubits based on bulk superconductors \cite{hyart2013flux} we consider system sizes $L\gg L_\gamma$.
Equation (\ref{eq:diff_sol}) suggests that $n_{\rm qp}(x)$ saturates at $n_{\rm max}= \gamma_{\rm br}n_{\rm CP}L^2/4D$ which can be understood as the density of broken Cooper pairs (in the absence of relaxation) created over the time it takes a QP to explore half of the system $L^2/4D$.
For large systems with $L>L_{\rm SC}$ this density will be cut off once it reaches $n_{\rm max}=n_{\rm SC}$ and the formerly neglected pair recombination processes become relevant.
Increasing the size of the system beyond $L_{\rm SC}$ will not change the density profile around the MZM but only add a larger region of density $n_{\rm SC}$ far away from the MZM.
We can therefore estimate the density $n_\gamma$ in the limit of a bulk superconductor by $n_\gamma(L=L_{\rm SC})$ which leads to poisoning rates of $\Gamma_{\gamma,\text{bulk}}=2\gamma_{\rm br}n_{\rm CP}L_{\rm SC}$ with $L_{\rm SC}=\sqrt{n_{\rm SC} D/n_{\rm CP}\gamma_{\rm br}}$. Using the estimates for $\gamma_{\rm br}$ based on Eq.~\eqref{eq:island_steady_state} one can rewrite $\Gamma_{\gamma,\text{bulk}}=\sqrt{2\gamma_0 D n_{\rm SC}^3/n_{\rm CP}}$ and with $n_{\rm SC}$ and $\tilde{\gamma}_0$ as quoted above we obtain poisoning times 
\begin{align}
\Gamma_{\gamma, \rm bulk}^{-1} \sim 0.1\mu {\rm s} \dots 1\mathrm{ms},\label{eq:bulk}
\end{align}
with corresponding $L_{\rm SC}\sim 1\dots 100$mm. Note that while the resulting estimated time scales $\Gamma_{\gamma,\text{bulk}}^{-1}$ are similar to existing literature estimates using bulk supercondcutors with constant densities $n_\text{SC}$ \cite{rainis2012majorana}, this  agreement is accidental as the underlying equations describe different physics \footnote{In Ref.~[6] the QPP rate is estimated by the tunneling rate of quasiparticles into the semiconductor assuming a bulk QP density of $n_{SC}$. We only consider a QP event as poisoning when QPs are entering a MZM and treat the rate of QPs tunneling between semiconductor and superconductor in Supplement~\ref{sup:coupled}, which leaves the results of the main text essentially unchanged.}.

A key advantage of our description is that it allows to model both the mesoscopic and bulk superconductors on the same footing and comparing Eqs.~\eqref{eq:meso} and~\eqref{eq:bulk} suggests that the poisoning times of Majorana qubits based on bulk superconductors are significantly worse than those of isolated mesoscopic islands. While the above estimates do not make it impossible to build Majorana qubits with bulk superconductors, the resulting coherence times will likely not be able to exceed those of conventional superconducting qubits. 

{\it Poisoning due to subgap states.---}We now turn to another type of decoherence possible in Majorana devices -- leakage of the parity from the MZM subspace into other subgap states.
This mechanism does not require non-equilibrium QPs. For simplicity we thus focus on \textit{equilibrium} QPP due to subgap states.
The concern here is that realistic Majorana devices may be disordered or inhomogeneous, and while the transport gap can be observed in the widely used transport experiments the (potentially zero) spectral gap is less accessible.
Thus equilibrium QPP is possible in topological phase if there are enough subgap states inside the wire that are close enough in position to the MZMs.
The equilibrium rate of a jump to a subgap state is:
\begin{align}\label{eq:vrh_rate}
\Gamma = \omega_a \exp \left( - \frac{2 x}{\xi} - \frac{\delta E}{k_B T} \right).
\end{align}
Here $x$ is the distance from the Majorana to the localization center of a subgap state, $\xi$ is the disordered coherence length, $\delta E$ is the energy difference to the target state, and $\omega_a$ is the ``attempt frequency" corresponding to the physical process that enables the tunneling event (for example, electron-phonon scattering, charge noise on gates \cite{Knapp18,aseev2018lifetime}, etc.).

As a specific scenario, we can take the average number of subgap states in a p-wave wire with finite mean free path $\ell$ arising from Gaussian disorder, as worked out in Ref.~\onlinecite{Brouwer2011}:
\begin{align}
\left< N(E) \right> \propto \frac{L}{\xi_0} \left( \frac{E}{\Delta_0} \right)^\eta, \quad \eta = 4\ell/\xi_0 - 2
\label{eq:NE}
\end{align}
where the proportionality constant is of order one, $\xi_0$ is the clean coherence length, $\Delta_0$ the gap in the zero-disorder limit, and the parameter $\eta$ describes the approach to a disorder-driven topological phase transition when $\ell \leq \xi_0 / 2$.
The \emph{apparent} coherence length, which governs the spatial overlap of Majorana wavefunctions and diverges at this transition, is given by $\xi^{-1} = \xi_0^{-1} - (2\ell)^{-1}$.
From Eq.~\eqref{eq:NE}, we can extract the energy window $[0,E']$, with $E' = \Delta_0 \left( \xi_0 / (2 x) \right)^{1/\eta}$, that contains on average $1$ state within a distance $x$ from the edge of the system.
Thus, to tunnel into a subgap state at distance $x$ an electron typically needs to gain an energy
\begin{align}
\delta E = \int_{0}^{E'} E \nu(E) dE = \frac{\Delta_0 \eta}{1 + \eta}
\left( \frac{\xi_0}{2 x} \right)^{1/\eta}.
\end{align}
The optimal tunneling distance is then obtained by maximizing the rate Eq.~(\ref{eq:vrh_rate}) with respect to $x$,
\begin{align}
x_{\rm opt} =
\frac{\xi}{2} \left( \frac{\eta + 2}{\eta} \right)^{\frac{\eta - 1}{\eta + 1}}
\left( \frac{1}{\eta + 1} \frac{\Delta}{k_B T} \right)^{\frac{\eta}{\eta+1}}.
\end{align}
Note that this expression has been written in terms of the disordered coherence length and transport gap, using $\xi / \xi_0 = (\eta + 2) / \eta$.
Then, finally, the typical leakage rate is obtained from Eq.~(\ref{eq:vrh_rate}) at $x_{\rm opt}$:
\begin{align}
\Gamma_{\rm sub} = \omega_a \exp \left[ - g(\eta)
	\left( \frac{\Delta}{k_B T} \right)^{\frac{\eta}{\eta + 1}} \right].
\end{align}
with $g(\eta) = \left( \eta+1 \right)^{\frac{1}{\eta+1}} \left( [\eta+2]/\eta \right)^{\frac{\eta-1}{\eta+1}}$.
Crucially, this rate is suppressed by a \emph{stretched} exponential in $\Delta/(k_B T)$.

\begin{figure}
	\includegraphics[width=0.95\columnwidth]{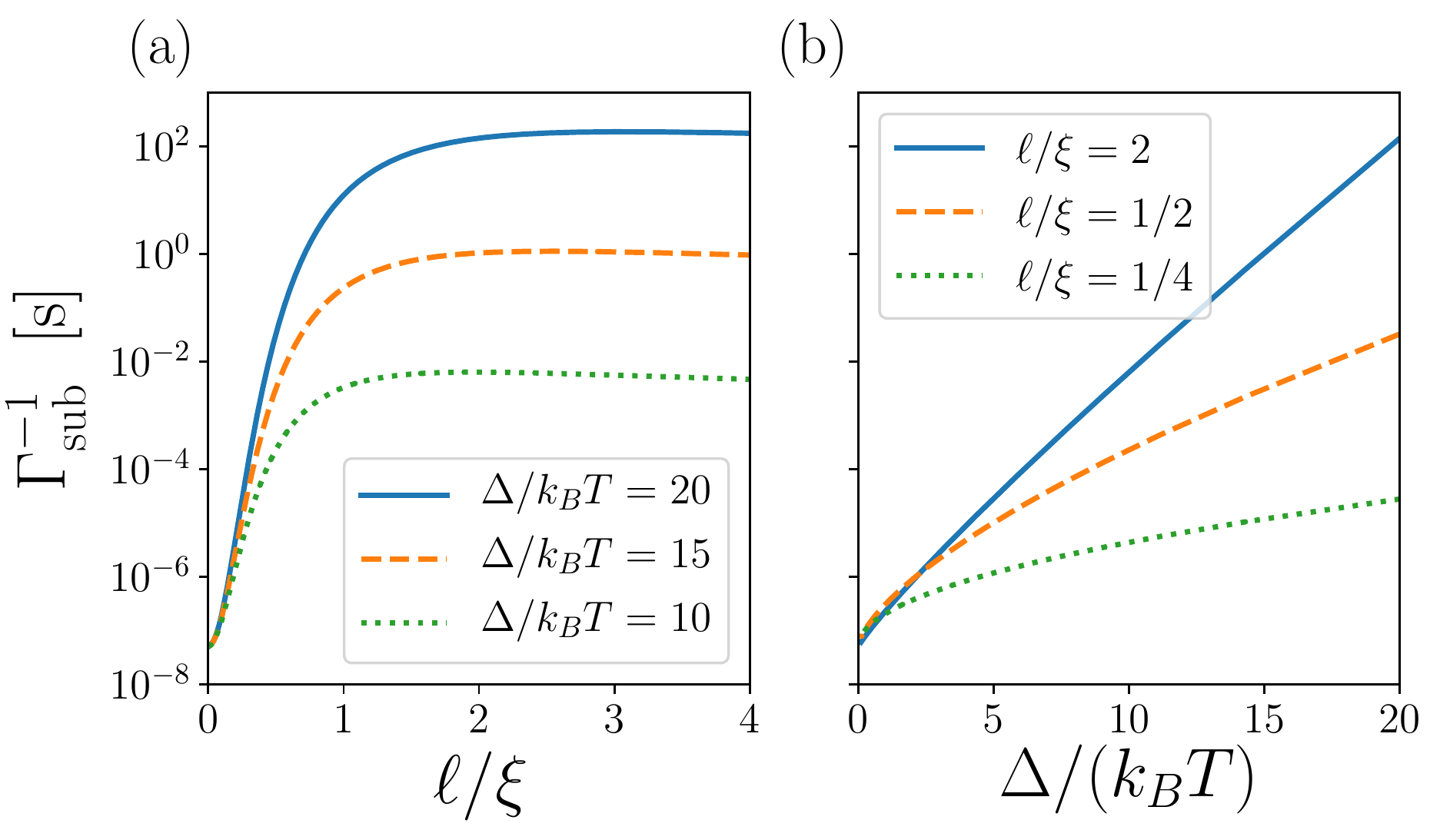}
	\caption{(a) Equilibrium poisoning time for transport gap $\Delta=100\mu$eV and several values of temperature as a function of $\ell/\xi$. (b) Stretched exponential temperature dependence of equilibrium poisoning time for several values of $\ell/\xi$.}
	\label{fig:3}
	\end{figure}

To gain intuition for the time scale in realistic devices we first take a (transport) topological gap of $\Delta \approx 100\mu$eV and a temperature $k_B T = 5\mu$eV ($\approx 50$mK).
We also assume the attempt frequency is the electron-phonon scattering rate $\omega_a=\gamma_0\sim (50 {\rm ns})^{-1}$.
For $\ell / \xi = 1/4$ (or $\ell / \xi_0 = 3/4$) we find a typical leakage rate of $\Gamma_{\rm sub}^{-1} \approx 30\mu$s, and already for $\ell / \xi = 1/2$ ($\ell / \xi_0 = 1$) we find $\Gamma_{\rm sub}^{-1} \approx 30$ms.
In Fig.~\ref{fig:3} we plot the leakage rate as a function of $\ell/\xi$ and of $\Delta/k_B T$.
It is apparent that at high temperatures or at the disorder-driven phase transition the poisoning time is given by the electron-phonon scattering, however the time can exceed seconds (with $\Delta / k_B T \approx 20$) for $\ell / \xi \gtrsim 1$ and this mechanism of QPP becomes essentially inoperative when $\ell / \xi \gtrsim 2$.
While these estimates are promising, they rely on a simple model which may be substantially modified in a real topological superconducting heterostructure, for example due to rare but strong impurities like dislocations, or the heterogeneous nature of the system \cite{liu2018impurity}.

In conclusion, we studied the effect of quasiparticle poisoning in Majorana qubits and found that non-equilibrium QPs are less harmful than expected from naive estimates based on the typical bulk quasiparticle concentration in conventional superconductors.
We have shown that poisoning due to non-equilibrium QPs is happening at the rate of the QP generation, which we expect to be of the order of seconds (or even days) and thus much slower than the timescales of qubit operations.
Another potential source of QPP is the presence of disorder-induced subgap states.
Our estimations show that Majorana parity leakage into such states is negligible for weak disorder. Only once the mean free path  becomes comparable to the coherence length, QPP may present a problem due to supgap states that proliferate as the system approaches the disorder-induced topological transition.

We acknowledge useful discussions with Chetan Nayak and Gijs de Lange.

\nocite{apsrev41Control}
\bibliographystyle{apsrev4-1}
\bibliography{qp}

\onecolumngrid
\appendix
\renewcommand{\appendixname}{Supplement}
\newpage
\section{Coupled diffusion equations}
\label{sup:coupled}
In the main text we discussed diffusion of quasiparticles in the parent superconductor ignoring the hybrid superconductor-semiconductor structure of typical one-dimensional topological superconductors. The semiconducting region might become relevant since it is generally described by a larger diffusion constant than the superconductor. The diffusion constant for early realizations of InAs nanowires has been experimentally estimated to be $D_{\rm w}\sim 10\mu{\text m}^2{\text s}^{-1}$ \cite{Jespersen09}. These nanowires were, however, of insufficient quality for topological superconductivity. Using $D_w=v_F^2 \tau$, in terms of the Fermi velocity $v_F$ and the scattering time $\tau$ we can estimate the diffusion constant for cleaner nanowires. A condition for topological superconductivity is $\tau> \hbar/\Delta$, where $\Delta$ is the topological gap (this is equivalent to the $\ell\lesssim\xi_0$ in the main text). Using $v_F>10^5 \text{m/s}$ and $\Delta\sim 100\mu\text{eV}$ we find a lower bound of $D_{\rm w}\gtrsim 100 \mu \text{m}^2/\text{ns}$ which is significantly larger than typical Al diffusion constants. For larger choices of Fermi velocity and less strong scattering $D_\text{w}$ can exceed the lower bound by several orders of magnitude.

Adding a channel with faster diffusion to the model of the main text will in general lead to more favorable estimates of the QP poisoning times since the effectiveness of the MZMs as quasiparticle traps is improved. To make theoretical progress in describing the hybrid system we consider the coupled diffusion equations
\begin{eqnarray}
	\dot{n}_{\rm qp,s} &=& 2\gamma_{\rm br}n_{\rm CP} +D\nabla^2 n_{\rm qp,s} - \sum_i\gamma_0 n_{\rm qp,s} f_\text{s}(\mathbf{x}-\mathbf{x}_i) - \tilde{\gamma}_0 n_{\rm qp,s}^2/n_{\rm CP} -\gamma_\text{s} n_\text{qp,s}+\gamma_\text{w}n_\text{qp,w} \label{eq:coupled1}\\
	\dot{n}_{\rm qp, w} &=& 2\gamma_{\rm br}n_{\rm w} +D_\text{w}\nabla^2 n_{\rm qp,w} - \sum_i\gamma_0 n_{\rm qp,w} f_\text{w}(\mathbf{x}-\mathbf{x}_i) - \tilde{\gamma}_0 n_{\rm qp,w}^2/n_{\rm w} -\gamma_\text{w} n_\text{qp,w}+\gamma_\text{s}n_\text{qp,s}\,,\label{eq:coupled2}
\end{eqnarray}
where we distinguish semiconductor nanowire variables and superconductor variables using the subscript ``w'' and ``s'', respectively. We denote the electron density in the nanowire as $n_\text{w}$, while $f_\text{w}$ and $f_\text{s}$ include the corresponding weight of the Majorana wavefunction in the superconductor and semiconductor, and we introduced the rate $\gamma_\text{s}$ ($\gamma_\text{w}$) for a QP transitioning from the superconductor to the nanowire (from the wire to the superconductor).

The transition rates can be estimated as $\gamma_\text{s} \sim g \Delta/ (\hbar n_\text{CP}V)$ and $\gamma_\text{w}\sim g v_F/L$ \cite{lutchyn2007,rainis2012majorana} in terms of the dimensionless conductance $g$ of the superconductor-semiconductor interface. Using $v_F=10^5 \text{m/s}$ and the values of $n_\text{CP}$ and the superconductor cross section assumed in the main text yields $\gamma_\text{w}/\gamma_\text{s}\sim 4\cdot 10^3$. The fact that $\gamma_\text{w}\gg\gamma_\text{s}$ reflects the substantial difference in the density of states of the metal and the semiconductor. As a consequence $n_\text{qp,w}\ll n_\text{qp,s}$ which allows to make simplifications to the coupled diffusion equations. First of all we can neglect the terms describing QP creation and recombination in the nanowire since almost all of the quasiparticles will be created and recombine (in the absense of MZMs) in the superconductor. In the absence of MZMs, the QP density in the nanowire is therefore largely controlled by the exchange of QPs with the superconductor. The strength of coupling between the semiconductor and superconductor is characterized by the relation between $\hbar\gamma_\text{w}$ and the parent gap $\Delta_\text{s}$ \cite{Stanescu11}, with the desired ``intermediate'' coupling when the two scales are equal \cite{antipov2018}. Therefore, except in the limit of very weak coupling, we can also neglect relaxation of QPs from the nanowire into MZM; since $\hbar\gamma_0 \ll \Delta_\text{s}$ QPs in the semiconductor are much more likely to transition into the superconductor (and possibly relax there) than directly relax into a MZM. Staying away from the regime of very weak coupling also ensures that there is no suppression of $f_\text{s}$ due to a small weight of the Majorana wave function in the superconductor.

With the above assumptions, and neglecting QP recombination we can solve Eqs.~\eqref{eq:coupled1},\eqref{eq:coupled2} in the same geometry as in the main text which yields
\begin{equation}
n_\text{qp,s}(x)= \left( 1+ \frac{\gamma_0 \xi}{D_\text{eff}}x\right)n_\gamma-\frac{\gamma_\text{br}n_\text{CP}}{D_\text{eff}}x^2+\gamma_\text{br}n_\text{CP}\left(\frac{1}{D_\text{s}}-\frac{1}{D_\text{eff}}\right)\lambda L \frac{\cosh(L/2\lambda)-\cosh((L-2x)/2\lambda)}{\sinh(L/2\lambda)}\label{eq:coupled_sol}	
\end{equation} 
where we only quoted the result for the density $n_\text{qp,s}$ since it is controlling relaxation into the MZMs. This solution introduces the effective diffusion constant $D_\text{eff}=D_\text{s}+\gamma_\text{s}D_\text{w}/\gamma_\text{w}$ due to the QPs spending a fraction $\gamma_\text{s}/\gamma_\text{w}$ of their time in the nanowire and a new length scale $\lambda$ given by $\lambda^{-2}=\gamma_\text{w}/D_\text{w}+\gamma_\text{s}/D_\text{s}$. The length scale $\lambda$ is roughly given by the minimum of the length scales $\sqrt{D_\text{w}/\gamma_\text{w}}$ and $\sqrt{D_\text{s}/\gamma_\text{s}}$ which describe the typical length a QP diffuses in the nanowire and superconductor before transitioning via the repsective rates $\gamma_\text{w},\gamma_\text{s}$. To get intuition about their typical values we use the intermediate coupling regime $\gamma_\text{w}\sim \Delta_\text{s}/\hbar$ in terms of the parent gap $\Delta_\text{s} = 200\mu\text{eV}$ and $D_\text{w}=100\mu \text{m}^2/\text{ns}$. We find $D_\text{s}/\gamma_\text{s}\sim 30\mu\text{m}$ and $D_\text{w}/\gamma_\text{w}\sim 0.3\mu\text{m}\sim \lambda$. Given the length scales considered in the main text we thus observe that the limit $\lambda \ll L$ is likely the most relevant.

We now discuss the similarities and differences of the solution \eqref{eq:coupled_sol} compared to Eq.~\eqref{eq:diff_sol} of the main text. Most importantly the density at $x=0$ and thus the rate of relaxation into the MZMs remains unchanged. The results of the main text regarding the Majorana poisoning scale thus remain unchanged as long as the approximations leading to solution \eqref{eq:coupled_sol} remain justified. The only effect of introducing the diffusion in the nanowire is thus a change in the shape of the density profile away from $x=0$. For $x\ll \lambda$ an expansion of the $\cosh$ and $\sinh$ terms reveals that $n_\text{qp,s}(x)$ is identical to the solution \eqref{eq:diff_sol} even for finite $x$. For $x\gg \lambda$ the density $n_\text{qp,s}(x)$ follows the shape of the main text solution with $D$ replaced by $D_\text{eff}$ up to a constant offset $\gamma_\text{br}n_\text{CP}\left(D_\text{s}^{-1}-D_\text{eff}^{-1}\right)\lambda L$. In the limit $\lambda \ll L$ the offset leads to only a minor shift of $n_\text{max}$. We thus conclude that the entire discussion in the main text already captures the most relevant mechanism for QP diffusion as long as the wire is not very weakly coupled to the supercondcutor. In the practically unlikely scenario of $\lambda \gtrsim L$ the discussion around the relevant saturation scale due to QP recombination needs to be adjusted taking into account a different value for $n_\text{max}$. 

\end{document}